\documentclass[aps,pre,showpacs,groupedaddress]{revtex4}%PRE
\usepackage{amsfonts}
\usepackage{mathrsfs}
\usepackage{graphicx}
\usepackage{amsmath}
\usepackage{amssymb}
\usepackage{amsbsy}

\begin{document}

\title{Message passing algorithms for the Hopfield network reconstruction: Threshold behavior and limitation}

\author{Haiping Huang}
%\email{hphuang@itp.ac.cn}
\affiliation{Key Laboratory of Frontiers in Theoretical Physics,
Institute of Theoretical Physics, Chinese Academy of Sciences,
Beijing 100190, China}
\date{\today}

\begin{abstract}
The Hopfield network is reconstructed as an inverse Ising problem by
passing messages. The applied susceptibility propagation algorithm
is shown to improve significantly on other mean-field-type methods
and extends well into the low temperature region. However, this
iterative algorithm is limited by the nature of the supplied data.
Its performance deteriorates as the data becomes highly magnetized,
and this method finally fails in the presence of the frozen type
data where at least two of its magnetizations are equal to one in
absolute value. On the other hand, a threshold behavior is observed
for the susceptibility propagation algorithm and the transition from
good reconstruction to poor one becomes sharper as the network size
increases.

\end{abstract}

\pacs{84.35.+i, 02.50.Tt, 75.10.Nr, 64.60.A-}
 \maketitle

%%%%%%%%%%%%%%%%%%%%%%%%%%%%%%%%%%%%%%%%%%%%%%%%%%%%%%%%%%%%%%%%%
\section{Introduction}
%%%%%%%%%%%%%%%%%%%%%%%%%%%%%%%%%%%%%%%%%%%%%%%%%%%%%%%%%%%%%%%%%
Message passing algorithms have important applications in various
contexts ranging from random constraint satisfaction
problems~\cite{Mezard-2002}, supervised
learning~\cite{Zecchina-2006} to information
theory~\cite{Richa-2001,Huang-2009}. In recent years, the idea of
message passing was introduced into the study of the inverse Ising
problem~\cite{Mezard-09,Aurell-2010,Marinari-2010}. Given observed
magnetizations (mean activities) and pairwise correlations
$\{m_{i},C_{ij}\}$, the inverse Ising problem aims at finding the
underlying parameters $\{h_{i},J_{ij}\}$, local fields and coupling
constants to describe the statistics of the experimental data which
can be collected either in real experiments (e.g., microarray
measurements in gene expression experiments or multi-electrode
recordings in a neuronal population) or in Monte Carlo simulations.
The active research of inverse Ising problem is mainly motivated by the
observation of correlated activity in the retinal
network~\cite{Shlens-2006,Nature-08,Shlens-2009}, the cortical
network~\cite{Tang-2008,Nature-10} and other biological
networks~\cite{Lezon-06,Seno-2008,Mora-2010}. Encouragingly, the
pairwise Ising model, as a least structured model, was shown to be
capable of capturing most of the correlation structure of the
network activity~\cite{Nature-06,Lezon-06,Tang-2008,Bialek-09ep}.
Based on the Ising model, computationally efficient inverse
algorithms were proposed to analyze multi-electrode recordings in
the salamander retina~\cite{Cocco-09} and to identify correlation
between amino acid positions in interacting
proteins~\cite{Weigt-2009}. The pairwise Ising model only requires
$\mathcal{O}(N^{2})$ parameters to describe the original
distribution and is thus attractive for dimensional reduction in
modeling vast amounts of biological data (for a general statistical
physics analysis of the pairwise Ising model on the inverse Ising
problem , see, e.g., Ref.~\cite{Roudi-2009b}).

For large system, the inverse Ising problem is known to be a hard
computational problem~\cite{Ack-1985,Hinton-1986,Hertz-1991}.
Various approximate schemes were proposed to tackle this problem. As
one of these approximations, message passing strategy looks
promising and the susceptibility propagation (SusProp) algorithm has
been derived to infer couplings of Sherrington-Kirkpatrick (SK)
model~\cite{Mezard-09}. SusProp solves a closed set of equations by
iteration, passing messages along the directed edges of the factor
graph representation~\cite{Frey-2001} of the problem. The new
message is computed only based on the incoming messages. This
locally updating feature makes SusProp fully distributed and
amenable to parallelization. Correlation information of any two
variables provided by SusProp can be used not only to decimation
procedure in solving some hard constraint satisfaction
problems~\cite{Higuchi-2010} but also to the network
reconstruction~\cite{Weigt-2009,Aurell-2010,Marinari-2010}.
Dependence of the performance of SusProp on the quality of the
originally observed data has been studied in
Ref.~\cite{Marinari-2010}. At high temperature, the quality of
reconstruction is constrained by the implementation precision of the
algorithm and the random noise embedded in the supplied data.
Statistical errors presented in the Monte Carlo noisy data could
also have detrimental effects on the reconstruction performance.
Aurell \textit{et al.} in a recent work~\cite{Aurell-2010} studied
the dynamical behavior of SusProp on inferring couplings from
synthetic data of SK model. They found that, at the low temperature
($T<4.0$), the algorithm doesn't converge typically with diverging
inferred couplings, however, by introducing a stopping criterion,
the threshold could be pushed to lower value. A transition from
reconstructible to non-reconstructible phase for SusProp was also
observed in their numerical simulations. High absolute magnetization
was claimed to have negative effects on the performance of the
algorithm. All aforementioned investigations were restricted to the
SK model. Mean field schemes based on inversion of correlation
matrix have been recently tested on Hopfield
networks~\cite{Huang-2010}. Regarding these mean field schemes, the
simple ones are naive mean field (nMF) method and independent pair
(ind) approximation, and the more advanced ones inversion of
Thouless-Anderson-Palmer (TAP) as well as Sessak-Monasson (SM)
approximation (for details, see Refs.~\cite{Roudi-2009,Huang-2010},
a brief description is also given in Appendix~\ref{app_mfs}). It was
shown that all mean field schemes fail to extract interactions
within a desired accuracy in the retrieval phase. Zhang \textit{et
al.}~\cite{Pan-2010} applied recently belief propagation plus an
auxiliary updating external field to infer couplings of the sparse
Hopfield network when the system settles in the retrieval phase.
They showed that inference error with sampling from single basin of
the stored pattern is much similar while error with sampling from
multiple basins is drastically reduced.

In the present work we will examine the reconstruction performance
of SusProp on both the fully connected and sparse Hopfield networks
and discuss the limitation of this message passing algorithm.
Improvements over other existing mean field methods are reported and
a threshold behavior relative to SusProp is observed in our
simulations on single instances. The rest of this paper is organized
as follows. The definition of the Hopfield network is given in
Sec.~\ref{sec_Hopf} followed by the detailed demonstration of
SusProp in Sec.~\ref{sec_MPA}. In Sec.~\ref{sec_result}, we report
improvements achieved by SusProp and discuss its limitation and
threshold behavior. Conclusions and future perspectives are devoted
to Sec.~\ref{sec_Con}.

%%%%%%%%%%%%%%%%%%%%%%%%%%%%%%%%%%%%%%%%%%%%%%%%%%%%%%%%%%%%%%%%%%%
\section{Hopfield Networks}
\label{sec_Hopf} The Hopfield model was proposed to mimic the memory
and recall functions of real neuronal
networks~\cite{Hopfield-1982,Amit-1989}. It yields rich statistical
physics properties and is moreover a simple model to assess the
efficiency of various inverse algorithms~\cite{Huang-2010}, which
may have some implications for neuroscience. The equilibrium
properties of the Hopfield network are governed by the following
Hamiltonian:
\begin{equation}\label{Hami}
    \mathcal{H}=-\sum_{i<j}J_{ij}\sigma_{i}\sigma_{j}
\end{equation}
where $\sigma_{i}$ describes the state of each neuron in the
network. $\sigma_{i}=+1$ indicates the spiking of neuron $i$ and
$\sigma_{i}=-1$ the silence of the neuron. Coupling $J_{ij}$ is
constructed according to the Hebb's rule:
\begin{equation}\label{Hebb}
     J_{ij}=\frac{1}{N}\sum_{\mu=1}^{P}\xi_{i}^{\mu}\xi_{j}^{\mu}
\end{equation}
where $\{\xi_{i}^{\mu}\}$ taking $\pm1$ with equal probability are
$P$ stored random patterns. The ratio of the number of stored
patterns to the network size $N$ is termed the memory load of the
network, i.e., $P=\alpha N$. In the fully connected network, each
neuron is connected to all the other neurons and no
self-interactions are assumed. The mean field behavior of the fully
connected Hopfield model has been thoroughly studied in
Refs.~\cite{Amit-198501,Amit-198502}. When $\alpha<0.138$,
paramagnetic, spin glass and metastable ferromagnetic retrieval
phases appear in order as the temperature decreases. The retrieval
phase becomes stable at low temperatures if $\alpha<0.051$. Replica
symmetry breaking occurs for the retrieval phase only at very low
temperatures. However, the replica symmetry solution for the spin
glass phase are unstable in the entire spin glass
phase~\cite{Toki-93}. For the sparse network, the coupling or
interaction is constructed as
\begin{equation}\label{J_spar}
    J_{ij}=\frac{l_{ij}}{l}\sum_{\mu=1}^{P}\xi_{i}^{\mu}\xi_{j}^{\mu}
\end{equation}
where $l$ is the mean degree of each neuron. In the thermodynamic
limit, $P$ scales as $P=\alpha l$ where $\alpha$ is the memory load.
No self-interactions are also assumed and the connectivity $l_{ij}$
follows the distribution:
\begin{equation}\label{distri}
    P(l_{ij})=\left(1-\frac{l}{N-1}\right)\delta(l_{ij})+\frac{l}{N-1}\delta(l_{ij}-1)
\end{equation}
Mean field properties of the sparse Hopfield network have been
discussed within replica symmetric approximation in
Refs.~\cite{Coolen-2003,Skantzos-2004}. Three phases (paramagnetic,
retrieval and spin glass phases) have also been observed in this
sparsely connected Hopfield network with arbitrary finite $l$. For
large $l$ (e.g., $l=10$), the phase diagram resembles closely that
of extremely diluted case~\cite{Watkin-1991,Canning-1992} where the
transition line between paramagnetic and retrieval phase is $T=1$
for $\alpha\leq 1$ and that between paramagnetic and spin glass
phase $T=\sqrt{\alpha}$ for $\alpha\geq 1$. The spin glass/retrieval
transition occurs at $\alpha=1$.

We simulate the Hopfield network using Glauber dynamics plus
simulated annealing techniques to collect enough experimental data
(totally $10^{4}$ samplings): ${m_{i}=\left<\sigma_{i}\right>_{{\rm
data}}, C_{ij}=\left<\sigma_{i}\sigma_{j}\right>_{{\rm
data}}-m_{i}m_{j}}$ where $\left<\cdots\right>_{{\rm data}}$ denotes
the average over the collected data (simulation details are given in
Appendix~\ref{app_sim}), then we estimate the couplings $\{J_{ij}\}$
between neurons based on these measured magnetizations and two-point
connected correlations, such that the resulting Ising distribution
$P_{{\rm Ising}}\propto \exp\left[
\sum_{i<j}J_{ij}^{*}\sigma_{i}\sigma_{j}\right]$ is able to provide
an accurate description of the statistics of the experimental data.
Note that $J_{ij}^{*}$ is the inferred value and has been scaled by
the inverse temperature $\beta$. The external field $h_{i}$ is zero
for all neurons in the current Hopfield model. In other
cases~\cite{Nature-06,Weigt-2009,Mora-2010}, the external field can
be used to represent the preferred direction of $\sigma_{i}$.

\section{Inferring couplings by passing messages}
\label{sec_MPA} Neurons in the Hopfield network usually interact
with each other to yield collective behavior at the network level.
$\{C_{ij}\}$ measure the tendency for each pair of neurons to spike
cooperatively. Together with the information about mean firing rates
$\{m_{i}\}$, they can be used as inputs to SusProp for the network
reconstruction. SusProp passes messages along the directed edges of
the network by iterative updating. To iterate SusProp, two kinds of
messages are needed. We first define the cavity magnetization
$m_{i\rightarrow j}$ as the message propagating from neuron $i$ to
neuron $j$, then define the other kind of message, namely the cavity
susceptibility $g_{i\rightarrow j,k}\equiv\frac{\partial
h_{i\rightarrow j}}{\partial h_{k}}$ where $h_{i\rightarrow j}$ is
termed cavity field of neuron $i$ in the absence of neuron $j$ and
$h_{k}$ is the local perturbation. The
SusProp then reads as follows:
\begin{widetext}
\begin{subequations}\label{SusP}
\begin{align}
m_{i\rightarrow j}&=\frac{m_{i}-m_{j\rightarrow i}\tanh J_{ij}}{1-m_{i}m_{j\rightarrow i}\tanh J_{ij}}\\
g_{i\rightarrow j,k}&=\delta_{ik}+\sum_{n\in \partial i\backslash
j}\frac{1-m_{n\rightarrow i}^{2}}{1-(m_{n\rightarrow i}\tanh
J_{ni})^{2}}
\tanh J_{ni}g_{n\rightarrow i,k}\\
J_{ij}^{{\rm new}}&=\epsilon\left[\frac{1}{2}\log\left(\frac{(1+\widetilde{C_{ij}})(1-m_{i\rightarrow j}m_{j\rightarrow i})}
{(1-\widetilde{C_{ij}})(1+m_{i\rightarrow j}m_{j\rightarrow i})}\right)\right]+(1-\epsilon)J_{ij}^{{\rm old}}\\
\widetilde{C_{ij}}&=\frac{C_{ij}-(1-m_{i}^{2})g_{i\rightarrow j,j}}{g_{j\rightarrow i,j}}+m_{i}m_{j}
\end{align}
\end{subequations}
\end{widetext}
where $\partial i\backslash j$ denotes neighbors of neuron $i$
except $j$, $\delta_{ik}$ is the Kronecker delta function and
$\epsilon$ serves as a damping factor. The damping factor allows the
new $J_{ij}$ to memorize a given fraction ($1-\epsilon$) of $J_{ij}$
computed at the last step (denoted as $J_{ij}^{{\rm old}}$), which
helps SusProp converge to a fixed point if the temperature is not
very low although the convergence is slowed down. In our current
simulations, we employ very small $\epsilon$ of order from
$\mathcal{O}(10^{-2})$ to $\mathcal{O}(10^{-4})$. Detailed
derivation of SusProp is given in Appendix~\ref{app_susp}. For other
discussions of this algorithm, we refer the reader to previous
works~\cite{Mezard-09,Higuchi-2010,Aurell-2010,Marinari-2010}. The
SusProp algorithm is able to estimate the correlation between any
two variables even if they are not directly linked in the
network~\cite{Higuchi-2010} and this information could be used
further to update couplings. Furthermore, the memory term $\epsilon$
ensures the update of $J_{ij}$ towards its true value step by step
when the temperature is not very low. For the fully connected
network, SusProp has the complexity of $\mathcal{O}(N^{3})$.

To assess the reconstruction performance of SusProp, we define
intuitively the inference error as
\begin{equation}\label{error}
    \Delta=\left[\frac{2}{N(N-1)}\sum_{i<j}(J_{ij}^{*}-J_{ij}^{{\rm true}})^{2}\right]^{1/2}
\end{equation}
where $J_{ij}^{*}$ is the inferred value of the coupling and
$J_{ij}^{{\rm true}}$ the original coupling constructed according to
the Hebb's rule Eq.~(\ref{Hebb}) or Eq.~(\ref{J_spar}). To run
SusProp, we initially set all $J_{ij}$ to be zero, and randomly
initialize for every edge of the network the message
$m_{i\rightarrow j}\in [-1.0,1.0]$ and $g_{i\rightarrow j,k}=0$ if
$i\neq k$ and $1.0$ otherwise. Then SusProp is iterated according to
Eq.~(\ref{SusP}) until the inferred couplings converge or the preset
maximal number of iterations $\mathcal{T}_{max}$ is saturated. In
our simulations, we adopt convergence criterion $\eta=10^{-4}$,
i.e., convergence of SusProp is identified once all updated
couplings have converged within precision $\eta$.

\begin{figure}
\centering
  \includegraphics[width=0.7\textwidth]{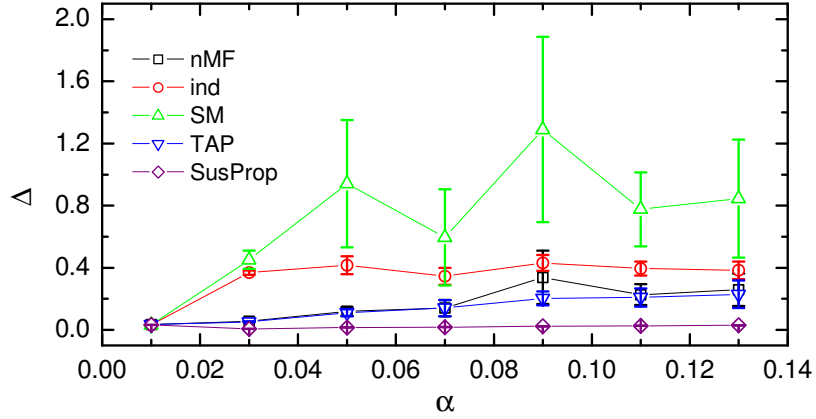}
\caption{(Color online) Comparison of reconstruction performances of
SusProp and other existing mean field schemes in the fully connected
Hopfield network. The inference error is plotted against the memory
load with $T=0.6, N=100$. Lines are guides to the eye. Each point is
an average over five random samples and error bars are also shown.
  }\label{Error_FVa}
\end{figure}

\begin{widetext}
\begin{center}
\begin{figure}
%\centering
    \includegraphics[bb=16 11 249 141,width=8.5cm]{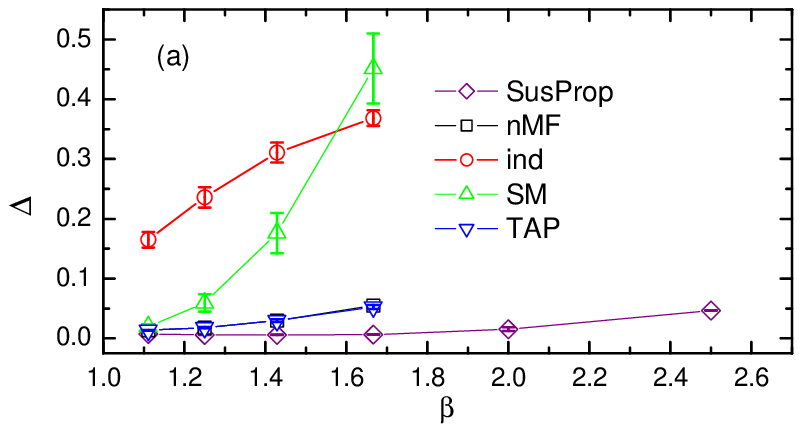}
    \hskip .5cm
    \includegraphics[bb=11 13 249 137,width=8.5cm]{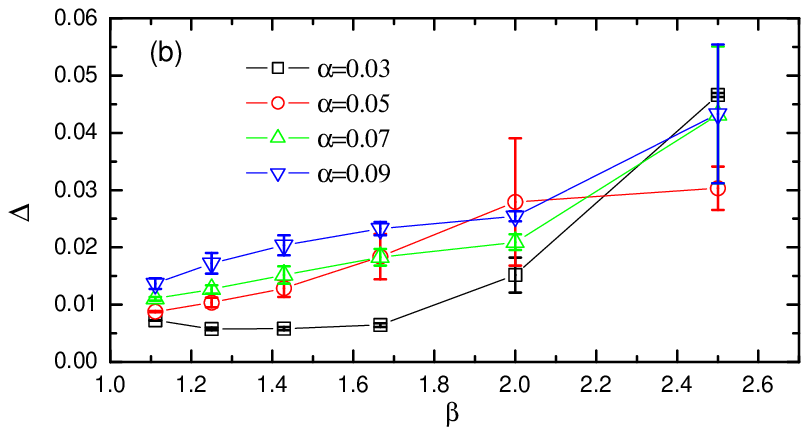}
    \vskip .2cm
  \caption{(Color online)
     The inference error versus temperature for the fully connected network with $N=100$.
     Lines are guides to the eye. Each point represents an average over five random samples and
     error bars are also shown. (a) Comparison of reconstruction performances of SusProp and other
     mean field schemes for $\alpha=0.03$. (b) Inference error versus temperature for SusProp with different
     memory loads.
  }\label{Error_FVT}
\end{figure}
\end{center}
\end{widetext}

%\begin{center}
\begin{figure}
\centering
    \includegraphics[width=0.7\textwidth]{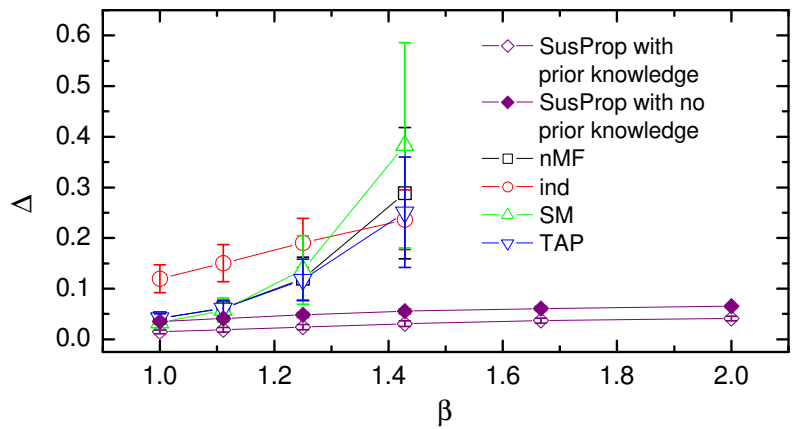}
  \caption{
    (Color online) Inference performances for the sparse Hopfield network with $\alpha=0.6, l=5, N=100$. Lines are
    guides to the eye. Each point is an average over five random samples and error bars are also shown. Efficiencies
    of SusProp and other mean field schemes are compared, so are the reconstruction performances of SusProp with and
    without prior knowledge on the sparseness of the network.
  }\label{Error_SVT}
\end{figure}
%\end{center}

\begin{figure}
\centering
    \includegraphics[width=0.7\textwidth]{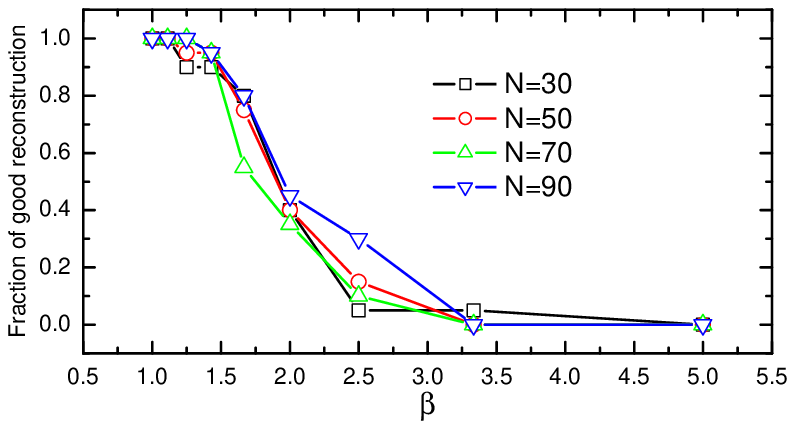}
  \caption{
    (Color online) Fraction of good reconstruction of fully connected Hopfield networks as a function of
    inverse temperature $\beta$ for SusProp. Different network sizes
    are considered with the same memory load $\alpha=0.1$. Twenty random samples are simulated and a good reconstruction is identified when $\Delta\lesssim 0.030$.
  }\label{Threshold}
\end{figure}

%%%%%%%%%%%%%%%%%%%%%%%%%%%%%%%%%%%%%%%%%%%%%%%%%%%%%%%%%%%%%%%%%%%%
\section{Reconstruction performances and Discussions}
\label{sec_result} To avoid the expensive computational cost, we
assess the reconstruction performance of SusProp only on small size
networks. The phase diagram of the Hopfield model has been derived
for the fully connected case~\cite{Amit-198501,Amit-198502} and the
finite connectivity case~\cite{Coolen-2003}. In the finite size
system, we distinguish different phases by two order parameters; one
is the overlap between the network configuration and the $\mu$th
pattern $m^{\mu}=\frac{1}{N}\sum_{i=1}^{N} \xi_{i}^{\mu}m_{i}$ and
the other is the mean-squared magnetization
$q=\frac{1}{N}\sum_{i=1}^{N}m_{i}^{2}$. To implement SusProp,
$\mathcal{T}_{max}$ is set to be $2000$ and we need to adopt an
appropriate damping factor to prevent the absolute updated $\tanh
J_{ij}$ from being larger than one.

Comparison of reconstruction performances of SusProp and other mean
field schemes is shown in Fig.~\ref{Error_FVa}. SusProp turns out to
be the most efficient, reducing the inference error by a significant
amount. Moreover, it seems to be less sensitive to the memory load
compared with other mean field methods. Fig.~\ref{Error_FVT} reports
the inference error as a function of temperature for various
reconstruction algorithms. SusProp operates fairly accurately and
extends the reconstructible region well into a much lower
temperature down to $0.4$. However, when the system settles in low
temperatures ($T<0.6$), SusProp first suffers highly magnetized data
and its performance gets worse with non-convergence, which may be
remedied by adopting much smaller $\epsilon$ and larger
$\mathcal{T}_{max}$ or by introducing a stopping
criterion~\cite{Aurell-2010}. When the temperature approaches lower
values (e.g., below $0.4$ in Fig.~\ref{Error_FVT} (a)), the
collected data will then become frozen, i.e., at least two of its
magnetizations equal one in absolute value, as a result, SusProp
yields diverging couplings and fails to reconstruct the network.
Actually, in the high absolute magnetization case, both
$|\widetilde{C_{ij}}|$ and $|m_{i\rightarrow j}|$ (and
$|m_{j\rightarrow i}|$) are very close to one on some edges $<ij>$.
We then rewrite $\widetilde{C_{ij}}$ as $\widetilde{C_{ij}}={\rm
sgn}(\widetilde{C_{ij}}) (1-\varepsilon_{ij})$, and similarly
$m_{i\rightarrow j}m_{j\rightarrow i}={\rm sgn}(m_{i\rightarrow
j}m_{j\rightarrow i})(1-\varepsilon_{ij,ji})$ where ${\rm
sgn}(\cdot)$ is a sign function; $\varepsilon_{ij}$ and
$\varepsilon_{ij,ji}$ are small positive
 values compared to one. Then one can readily recast $\tanh
J_{ij}$ according to Eq.~(\ref{SusP}) as
\begin{equation}\label{diver}
    \tanh J_{ij}=\frac{{\rm sgn}(\widetilde{C_{ij}})\varepsilon_{ij,ji}-{\rm sgn}(m_{i\rightarrow j}m_{j\rightarrow i})\varepsilon_{ij}}
    {\varepsilon_{ij,ji}+\varepsilon_{ij}}
\end{equation}
for $\widetilde{C_{ij}}m_{i\rightarrow j}m_{j\rightarrow i}>0$ and
$|\tanh J_{ij}|=1.0$ for $\widetilde{C_{ij}}m_{i\rightarrow j}
m_{j\rightarrow i}<0$. If $\widetilde{C_{ij}}m_{i\rightarrow
j}m_{j\rightarrow i}>0$ and neither of $\varepsilon_{ij}$ and
$\varepsilon_{ij,ji}$ vanishes, the estimated $J_{ij}$ remains
finite and SusProp does work. In other cases (e.g.,
$|\widetilde{C_{ij}}|=1.0$ and $|m_{i\rightarrow j}m_{j\rightarrow
i}|\neq1.0$), $J_{ij}$ suffers divergence and SusProp is unable to
infer the couplings. In this situation, at least two of the supplied
magnetizations are equal to one in absolute value, e.g.,
$|m_{i}|=1.0$ and $|m_{j}|=1.0$, under the update rule
Eq.~(\ref{SusP}), $m_{i\rightarrow j}={\rm sgn}(m_{i})$ and
$m_{j\rightarrow i}={\rm sgn}(m_{j})$. These highly polarized
messages will then spread out in the network, which yields an
infinite value for $J_{ij}$ as explained above. A simple physical
interpretation is that, in this case, $\widetilde{C_{ij}}\simeq
m_{i}m_{j}$, and this implies that some neurons in the network tend
to behave independently of other neurons and thus SusProp couldn't
get all information about correlations of the network, which leads
to the failure of reconstruction. In the current Hopfield model, the
frozen type data does appear for small enough $\alpha$ and $T$ where
the system gets trapped by one of stable or metastable memory
states~\cite{Huang-2010}. Increasing $\alpha$ but maintaining low
$T$, many spurious minima show up and Monte Carlo sampling becomes
very difficult~\cite{Toki-93,Lenka-PHD08}, which produces fairly
noisy collected data. On the other hand, the current SusProp hasn't
taken into account the complex structure of the phase space,
therefore it remains a non-trivial issue for SusProp to deal with
this more involved case. The reconstruction error against
temperature is also shown with respect to different memory loads for
SusProp in Fig.~\ref{Error_FVT} (b). In the high temperature region
($T>0.6$), the smaller the memory load is, the more precisely
SusProp reconstructs the Hopfield network. As temperature decreases
further, SusProp becomes less precise for all memory loads while
still maintaining a relatively small error.

SusProp is applied to reconstruct the fully connected Hopfield
network, whereas, it shows a surprisingly good performance. If the
network is sparse and locally treelike, SusProp is believed to be
fast and able to give a precise estimation. To test its efficiency
on reconstructing the sparse network, we compare performances of
SusProp with those of other mean field schemes in
Fig.~\ref{Error_SVT}. It is clearly shown that SusProp performs
exceedingly well particularly in the low temperature region (down to
$T=0.5$) and exhibits less sample-to-sample fluctuations. If we have
a prior knowledge of the sparseness of the network, i.e., we know a
priori the connectivity pattern of the network, the inference error
could be reduced substantially. In this case, we only infer the
strength of interaction between neurons which are really connected.
In fact, when the system is presented at the high temperature, one
can reconstruct the network using an appropriate cutoff since the
estimated couplings between unconnected neurons are very small
compared to those between really connected neurons.

In Fig.~\ref{Threshold}, we report the fraction of good
reconstruction versus temperature for different network sizes at
fixed memory load $\alpha=0.1$. A good reconstruction is identified
when $\Delta\lesssim0.030$ and SusProp converges within
$\mathcal{T}_{max}$. As increasing temperature, a threshold behavior
is observed and the transition becomes sharper with growing network
size. Given large enough $N$, the probability that SusProp gives
good reconstruction tends to be zero when the temperature is below
the critical value, and at a high enough temperature, SusProp
succeeds in reconstructing the network in all instances. The
critical temperature is estimated to be about $0.6$. For a more
precise estimation, more samples and larger network size are
required. The threshold behavior of SusProp on network
reconstruction has also been observed in SK
model~\cite{Aurell-2010}.

%%%%%%%%%%%%%%%%%%%%%%%%%%%%%%%%%%%%%%%%%%%%%%%%%%%%%%%%%%%%%%%%%%%%%%%
\section{Conclusion and outlook}
\label{sec_Con}
SusProp solves the inverse Ising problem by
iteratively updating messages along the directed edges of the
network, and is shown in the present work to outperform all other
mean-field-type schemes and extend the reconstructible region into
lower temperatures. We also study the fraction of good
reconstruction as a function of temperature and a threshold behavior
is observed. The transition from good reconstruction to poor one
becomes sharp as increasing network size and the critical
temperature is estimated to be about $0.6$. SusProp is also
amazingly efficient for reconstructing sparse Hopfield network and
its performance can be further improved by introducing a prior
knowledge about the sparseness of the network. The sparse case is
more relevant in modeling real biological data than its dense
counterpart. We hope our analysis of SusProp on the Hopfield network
reconstruction can be extended to the more biologically relevant
cases.

At high temperatures, the performance of SusProp is believed to be
limited by the quality of the supplied data~\cite{Marinari-2010}.
Once the system is presented at the low temperature, the efficiency
of SusProp is also determined by the nature of the input data. In
the presence of highly magnetized data, the reconstruction
performance of SusProp gets deteriorated with a high inference
error. Furthermore, the frozen type data makes updated couplings
diverge and any value of damping factor can not overcome this
hurdle, reminiscent of the fact that the frozen phase in random
constraint satisfaction problems is most difficult for any known
algorithm~\cite{Lenka-08,Lenka-PHD08}.

For large enough $\alpha$ but low enough $T$, the system enters the
spin glass phase where Glauber dynamics is easily trapped by one of
the spurious minima correlated or uncorrelated with the stored
patterns. SusProp fails to extract couplings precisely in this
region since it does not take into account the complex structure of
the phase space. In fact, at finite temperatures, the support of
cavity field distributions becomes real-valued and a sampling
procedure is required. On the other hand, other values of Parisi
parameter (smaller than the optimal value associated with the ground
states) also carry physical information and can be used to describe
the metastable states which trap Glauber dynamics~\cite{Mezard-04}.
However, searching for an optimal Parisi parameter (also known as
replica symmetry breaking parameter) is also a time consuming task
for the network reconstruction~\cite{Kappen-2006}. Rather, provided
that Glauber dynamics gets stuck in some metastable state for a very
long time, how much information can be extracted from this state
about the couplings of the network remains an important issue for
future study.

\section*{Acknowledgments}

%\begin{acknowledgments}
Helpful discussions with Erik Aurell, Haijun Zhou and Pan Zhang are
acknowledged. The present work was in part supported by the National Science Foundation
of China (Grant numbers 10774150 and 10834014) and the China
973-Program (Grant number 2007CB935903).
%\end{acknowledgments}
\appendix
\section{Mean field schemes for network reconstruction}
\label{app_mfs}
\subsection{Naive Mean-Field Method}
\label{subsec:nMF}
 The naive mean field theory gives
$m_{i}=\tanh\left(h_{i}+\sum_{k\neq i}J_{ik}m_{k}\right)$ where
$h_{i}$ is the external field and $m_{i}=\left<\sigma_{i}\right>$
the magnetization. Using the fluctuation-response relation,
\begin{equation}\label{corre}
    C_{ij}=\frac{\partial m_{i}}{\partial
    h_{j}}=(1-m_{i}^{2})\left[\delta_{ij}+\sum_{k\neq i}J_{ik}C_{kj}\right]
\end{equation}
one obtains the nMF prediction of couplings,
\begin{equation}\label{nmf}
    J_{ij}^{{\rm nMF}}=(\mathbf{P}^{-1})_{ij}-(\mathbf{C}^{-1})_{ij}
\end{equation}
where $\mathbf{P}_{ij}=(1-m_{i}^{2})\delta_{ij}$.
\subsection{Independent-Pair Approximation}
\label{subsec:Ind}

In this approximate scheme, each pair of neurons are independent of
other neurons of the system, i.e., their joint probability
$P(\sigma_{i},\sigma_{j})\propto\exp\left[h_{i}^{(j)}\sigma_{i}+h_{j}^{(i)}\sigma_{j}+J_{ij}\sigma_{i}\sigma_{j}\right]$
where $h_{i}^{(j)}(h_{j}^{(i)})$ is the local field neuron $i(j)$
feels when neuron $j(i)$ is removed from the system. Then the ind
prediction is given by
\begin{equation}\label{ind}
    J_{ij}^{{\rm ind}}=\frac{1}{4}\log\left[\frac{(1+C_{ij}^{'})^{2}-(m_{i}+m_{j})^{2}}{(1-C_{ij}^{'})^{2}-(m_{i}-m_{j})^{2}}\right]
\end{equation}
where $C_{ij}^{'}=C_{ij}+m_{i}m_{j}$.
\subsection{Sessak-Monasson Approximation}
\label{subsec:SM}

The SM prediction of couplings is derived based on a perturbative
expansion in the correlations~\cite{SM-09} and it can be formulated
as
\begin{equation}\label{SM}
    J_{ij}^{{\rm SM}}=J_{ij}^{{\rm nMF}}+J_{ij}^{{\rm ind}}-\frac{C_{ij}}{(1-m_{i}^{2})(1-m_{j}^{2})-C_{ij}^{2}}
\end{equation}
\subsection{Inversion of TAP Equations}
\label{subsec:InvTAP}

The usual TAP equation reads $h_{i}=\tanh^{-1}m_{i}-\sum_{j\neq
i}J_{ij}m_{j}+m_{i}\sum_{j\neq
i}J_{ij}^{2}(1-m_{j}^{2})$~\cite{Mezard-1987}. Differentiating the
field $h_{i}$ with respect to the magnetization $m_{j}$, one readily
obtains the TAP prediction equation,
\begin{equation}\label{TAP}
    (\mathbf{C}^{-1})_{ij}=\frac{\partial h_{i}}{\partial m_{j}}=-J_{ij}^{{\rm
    TAP}}-2(J_{ij}^{{\rm TAP}})^{2}m_{i}m_{j}
\end{equation}
\section{Simulation details}
\label{app_sim}

The rule for Glauber dynamics can be generally expressed as
$P(\sigma_{i}\rightarrow-\sigma_{i})=\frac{1}{1+\exp(\beta\Delta\mathcal{H}_{i})}$
where $\Delta\mathcal{H}_{i}$ is the energy change due to such a
flip. For the current Hopfield model, the dynamics rule is recast
into
\begin{equation}\label{GDrule}
    P(\sigma_{i}\rightarrow-\sigma_{i})=\frac{1}{2}\left[1-\sigma_{i}\tanh\beta h_{i}\right]
\end{equation}
where  $\beta$ is the inverse temperature and $h_{i}=\sum_{j\neq
i}J_{ij}\sigma_{j}$ is the local field acting on $\sigma_{i}$.

In our numerical simulations, we update the state of each neuron
according to Eq.~\ref{GDrule} in a randomly asynchronous manner. We
define a Glauber dynamics step as $N$ proposed flips. Introducing
simulated annealing strategy, we set the initial temperature to be
$1.0$ and the cooling rate $0.005$. At each intermediate
temperature, we run $10^{4}$ Glauber dynamics steps. When the
temperature is decreased to the desired one, we run another $2\times
10^{6}$ steps to calculate magnetizations and correlations. We
sample the state of the network every $200$ steps. For high
temperatures ($\geq 1.0$), we run totally $4\times 10^{6}$ steps,
among which the first $2\times 10^{6}$ steps are run for the system
to reach the equilibrium state and the other $2\times 10^{6}$ steps
for calculating magnetizations and correlations.
\section{Derivation of SusProp update rules}
\label{app_susp}

To derive Eq.~\ref{SusP}, we first derive the susceptibility
propagation equations for general $K$-body interaction problem
($K=2$ for the Hopfield model). Using the factor graph
representation~\cite{Frey-2001}, we denote $a$ as the function node
representing the constraint imposed on a subset of spins
$\boldsymbol{\sigma}_{\partial a}$ ($\partial a$ denotes
neighbors of the function node $a$), and $i$ as variable node
representing the spin on the factor graph. The belief propagation is then
formulated as~\cite{Huang-2009}
\begin{subequations}\label{BPeq}
\begin{align}
  \begin{split}
  m_{i\rightarrow a}&\equiv \tanh h_{i\rightarrow a}
=\tanh\left(\sum_{b\in\partial i\backslash a}u_{b\rightarrow
i}\right)=
\tanh\left[\tanh^{-1}(m_{i})-u_{a\rightarrow i}\right]\\
&=\frac{m_{i}-\tanh J_{a}\prod_{j\in\partial a\backslash
i}m_{j\rightarrow a}} {1-m_{i}\tanh J_{a}\prod_{j\in\partial
a\backslash i}m_{j\rightarrow a}}
\end{split}\\
\tanh u_{b\rightarrow i}&=\tanh J_{b}\prod_{j\in\partial b\backslash
i}m_{j\rightarrow b}
  \end{align}
\end{subequations}
where $h_{i\rightarrow a}$ is the cavity field (correspondingly $m_{i\rightarrow a}$ is the cavity magnetization) acting on spin
$\sigma_{i}$ in the absence of $a$; $u_{b\rightarrow i}$ the cavity
bias when $i$ is involved in $b$ only, and $h_{i\rightarrow a},
u_{b\rightarrow i}$ as well as $J_{b}$ have been rescaled by $\beta$.

We define cavity susceptibility $g_{i\rightarrow
a,k}\equiv\frac{\partial h_{i\rightarrow a}}{\partial h_{k}}$. From
the belief propagation equations, one readily gets the update rule
for $g_{i\rightarrow a,k}$:
\begin{equation}\label{Cavity_sus}
   \begin{split}
  g_{i\rightarrow a,k}&=\delta_{ik}+\sum_{b\in\partial i\backslash a}\frac{\partial u_{b\rightarrow i}}{\partial
h_{k}}\\
&=\delta_{ik}+\sum_{b\in\partial i\backslash a}\frac{\tanh
J_{b}}{1-\left(\tanh J_{b}\prod_{j\in\partial b\backslash
i}m_{j\rightarrow b}\right)^{2}}\sum_{j\in\partial b\backslash
i}\left[\prod_{n\in\partial b\backslash i,j}m_{n\rightarrow
b}\right]g_{j\rightarrow b,k}(1-m_{j\rightarrow b}^{2})
\end{split}
\end{equation}
Using the identity~\cite{Higuchi-2010}:
\begin{equation}\label{identity}
\frac{\partial u_{a\rightarrow i}\left(\{h_{j\rightarrow
a}\}_{j\in\partial a\backslash i}\right)}{\partial{h_{n\rightarrow
a}}}
=\frac{\left<\sigma_{n}\sigma_{i}\right>-\left<\sigma_{n}\right>\left<\sigma_{i}\right>}{1-\left<\sigma_{i}\right>^{2}}
\end{equation}
where $i,n\in\partial a$ and $\left<\cdots\right>$ denotes the
average under the joint probability distribution ${\rm
Prob}(\boldsymbol{\sigma}_{\partial a})$ which can be computed from
the belief propagation equations, one can re-express the
correlations $C_{ij}$ through the fluctuation-response relation:
\begin{equation}\label{correlation}
   \begin{split}
  C_{ij}&=\frac{\partial m_{i}}{\partial h_{j}}=(1-m_{i}^{2})\left[\delta_{ij}+\sum_{b\in\partial i}\frac{\partial u_{b\rightarrow i}}
{\partial h_{j}}\right]\\
&=(1-m_{i}^{2})\left[g_{i\rightarrow a,j}+\sum_{n\in\partial
a\backslash i}\frac{\partial u_{a\rightarrow i}}{\partial
h_{n\rightarrow a}}\frac{\partial h_{n\rightarrow a}}{\partial
h_{j}}\right]\\
&=(1-m_{i}^{2})\left[g_{i\rightarrow a,j}+\sum_{n\in\partial
a\backslash i}\frac{\partial u_{a\rightarrow i}}{\partial
h_{n\rightarrow a}}g_{n\rightarrow a,j}\right]
\end{split}
\end{equation}
For the simple case, the Hopfield network, $J_{a}=J_{ij}$, and the
constraint $a$ involves only two neurons, say $i$ and $j$. To obtain
the update rule for $J_{ij}$, we compute
$\widetilde{C_{ij}}\equiv\left<\sigma_{i}\sigma_{j}\right>$ directly
by assuming ${\rm Prob}(\sigma_{i},\sigma_{j})\propto
\exp\left(J_{ij}\sigma_{i}\sigma_{j}+h_{i\rightarrow
j}\sigma_{i}+h_{j\rightarrow i}\sigma_{j}\right)$, finally we get
\begin{equation}\label{coupling}
    \tanh J_{ij}=\frac{\widetilde{C_{ij}}-m_{i\rightarrow j}m_{j\rightarrow i}}
{1-\widetilde{C_{ij}}m_{i\rightarrow j}m_{j\rightarrow i}}
\end{equation}
where $\widetilde{C_{ij}}$ can be evaluated from Eqs.~\ref{identity}
and~\ref{correlation}, i.e.,
\begin{equation}\label{tide_cij}
\widetilde{C_{ij}}=\frac{C_{ij}-(1-m_{i}^{2})g_{i\rightarrow
j,j}}{g_{j\rightarrow i,j}}+m_{i}m_{j}
\end{equation}
From Eqs.~\ref{BPeq} and ~\ref{Cavity_sus}, the update rules for
$m_{i\rightarrow j}$ and $g_{i\rightarrow j,k}$ are finally obtained
as follows,
\begin{subequations}\label{SusProp}
\begin{align}
m_{i\rightarrow j}&=\frac{m_{i}-m_{j\rightarrow i}\tanh J_{ij}}{1-m_{i}m_{j\rightarrow i}\tanh J_{ij}}\\
g_{i\rightarrow j,k}&=\delta_{ik}+\sum_{n\in \partial i\backslash
j}\frac{1-m_{n\rightarrow i}^{2}}{1-(m_{n\rightarrow i}\tanh
J_{ni})^{2}} \tanh J_{ni}g_{n\rightarrow i,k}
\end{align}
\end{subequations}
Introducing additionally a damping factor $\epsilon$,
Eqs.~\ref{coupling},~\ref{tide_cij} and~\ref{SusProp} are the very
SusProp we have presented in Sec.~\ref{sec_MPA}.

%%%%%%%%%%%%%%%%%%%%%%%%%%%%%%%%%%%%%%%%%%%%%%%%%%%%%%%%%%%%%%%%%%%%%
%\bibliography{ref}

%%%%%%%%%%%%%%%%%%%%%%%%%%%%%%%%%%%%%%%%%%%%%%%%%%%%%%%%%%%%%%%%%%%%%

\end{document}